\documentstyle[aps,prl,epsf]{revtex}
\begin{document}
\twocolumn[\hsize\textwidth\columnwidth\hsize\csname@twocolumnfalse%
\endcsname

\title{Spinor Bosonic Atoms in Optical Lattices:\\
Symmetry Breaking and Fractionalization}
\author{Eugene Demler$^1$ and Fei Zhou $^{2,3}$}
\address{$^1$ Physics Department, Harvard University, Cambridge, MA 02138}
\address{$^2$ NECI, 4 independence way, Princeton, NJ 08540, USA}
\address{$^3$ ITP, University of Utrecht, Princetonplein 5, 
3584 CC Utrecht, The Netherlands\footnote{Permanent address}}

\date{\today}

\maketitle

\begin{abstract}
We study  superfluid and Mott insulator phases
of cold spin-1 Bose atoms with antiferromagnetic interactions
in an optical lattice,
including a  usual polar condensate phase, a condensate of
singlet pairs, a crystal spin nematic phase, and a spin
singlet crystal phase.  We suggest a possibility of
exotic fractionalized phases of spinor BEC 
and discuss them in the language of topological defect condensation
and $Z_2$ lattice gauge theory.

\end{abstract}
 ]

Recent experiments on the condensation of Bose gases with internal
degrees of freedom initiated a considerable amount of work on
understanding the nature of spinor BEC. In particular Bose
condensation in alkali atoms with nuclear spin $I=3/2$ that have three
low energy hyperfine states and therefore behave as $F=1$ bosons, have
been a subject of active experimental 
\cite{Stamper98,Stenger98,Ketterle00,nature} and theoretical 
research\cite{Ho98,Ohmi98,Law98,Ho00,Castin00,Zhou99}. 
It was pointed out
that the ground state of atoms with antiferromagnetic interactions in
an optical trap is not a usual condensate of a macroscopic number of atoms in
a single quantum state, but a complicated many body state of atoms
arranged into a total singlet\cite{Law98,Ho00,Castin00,Zhou99}.
%The rich physics of spinor Bose gases stems from the fact that
%condensation of individual atoms  breaks not only the charge symmetry,
%related to the total number of particles, but also a spin symmetry of
%rotations between the internal states of atoms. 
In this article we
examine  the problem of Bose  $F=1$ atoms with
antiferromagnetic interaction (spin symmetric interactions are assumed
throughout the paper) in optical lattices - arrays of microscopic
potentials created by interfering laser beams
\cite{Grynberg,Haman98,Friebel98,Rayzen97,Guidoni97,Petsas94,Deutsch98,Han00,Kasevich}.
The dynamics of spinless bosonic atoms in arrays of optical wells
may be described by  a Bose-Hubbard model with the possibility
of quantum phase transitions between 
insulating and superfluid phases induced by varying the properties
of the laser light \cite{Jaksch98}.
Here, we propose several new phases that will appear 
for $F=1$ bosons due to 
an interplay between spin and charge degrees of freedom: singlet
pair condensate that only breaks charge symmetry, a spin nematic
crystal phase that only  breaks spin symmetry, and a novel ``strong
coupling pairing phase'' that breaks both spin and charge symmetry but
is  distinct from a simple polar BEC. 
These phases may be
considered as ``fragmented condensates'' \cite{Nozieres82},
and have fractionalized topological excitations: half-vortices 
and $\pi$-disclinations.
Josephson type experiments may be used to distinguish among
insulating and various superfluid phases.
We also conjecture the possibility of
fractionalized phases in spinor Bose gases.
Fractionalization is
characterized by a topological order and may coexist with any broken
continuous symmetry \cite{Senthil99}.

In what follows we consider the  case  of antiferromagnetic
interaction between the atoms $ a_0 > a_2$, where 
$a_F$ is an $s$-wave scattering length in the $F$ channel.
In the individual wells we take wavefunctions that are superpositions
of the mean-field solutions of the interacting problem
in the absence of tunneling between the wells
\cite{Castin00,Zhou99}
$
 |\psi \rangle_N $=$ \int_{\bf n} \psi({\bf n}) |N, {\bf n} \rangle
$, with
$
|N, {\bf n} \rangle = \frac{1}{\cal N}( n_x a_x^{\dagger} + n_y
a_y^{\dagger} 
+ n_z a_z^{\dagger} )^N |0\rangle
$. Here ${\bf n}$ is a real unit vector, $a_\alpha^\dagger$ are boson
creation operators in the lowest vibrational state, 
$|0\rangle $ is the vacuum state, and ${\cal N}$ is a 
normalization factor.   Two properties of
these wavefunctions are important: (i) they satisfy
a symmetry condition $|N, -{\bf n} \rangle = (-)^N |N, {\bf n}
\rangle$ \cite{Zhou99};
(ii) in the limit of large $N$ states that correspond to
different ${\bf n}$'s are orthogonal to each other.
Hamiltonian
of an array of identical optical wells with spin symmetric
tunneling is given by 
\cite{long,disclaimerNlarge}
\begin{eqnarray}
{\cal H} &=& \sum_i {\cal H}_i +\sum_{ij} {\cal H}_{ij}
\nonumber\\
{\cal H}_i &=&  \frac{u}{2}\,  N_i^2 - \mu N_i 
+\frac{g}{2}\, {\bf L}_i^2
\nonumber\\
{\cal H}_{ij} &=& - 2 t\, {\bf n}_i {\bf n}_j\,\, ( b_i^\dagger b_j
+  b_j^\dagger b_i )
\label{Harray}
\end{eqnarray}
where we defined the charge creation and annihilation 
operators that  change the number of particles $N_i$, but not the direction
of ${\bf n}_i$: 
$ b_i^{\dagger} |N_i, {\bf n}_i \rangle = (N_i+1)^{1/2} |N_i+1, {\bf n}_i
\rangle$; the  number of particles in each well $ N_i = b_i^{\dagger} b_i$;
and the angular momentum operators ${\bf L}_i = 
-\,i\, {\bf n}_i \times \frac{\partial}{\partial {\bf n}_i}$
that describe a collective spin in well $i$ \cite{Law98,Zhou99}. 
Parity condition on the wavefunctions implies constraint
$N_i+L_i=$even for all $i$, and
for simplicity in (\ref{Harray}) we chose to work in the
grand canonical ensemble using a chemical potential $\mu$.
%%%%%%%%%%
For a single well Hamiltonian (\ref{Harray})
correctly gives a spin singlet ground state
if the number of particles is even
\cite{Law98,Ho00,Castin00,Zhou99}.
%%%%%%%%%%%%
When optical lattices are produced by  lasers with
wavelength $\lambda$, they  create an optical potential 
$V({\bf x})= V_0 \sum_i \sin^2(k x_i)$. We can estimate
parameters of the Hamiltonian (\ref{Harray}) as
$
g=\frac{2 \pi^2}{3}E_R \frac{(a_0-a_2)}{\lambda} (\frac{V_0}{E_R})^{3/4}
$, 
$
U=\frac{2\pi^2}{3}E_R \frac{(a_0+2a_2)}{\lambda} (\frac{V_0}{E_R})^{3/4}
$,
$
t= (4 V_0 E_R)^{1/2} Exp\{-2\pi (\frac{V_0}{E_R})^{1/2}
(1-2 (\frac{E_R}{V_0})^{1/2})\}
$, 
where  $E_R=\hbar^2k^2/2M$ is the recoil
energy.  
The average number of particles in each well
will be related to the atomic gas density  $n$ as  $\bar{N}=
n\, ( \frac{\lambda}{2\pi})^3 \,(\frac{E_R}{V_0})^{3/4}$.
The obvious parameters
to tune in experiments is $\bar{N}$ 
(e.g. by varying boson density) and the strength
of periodic potential $V_0$ (by tuning the laser intensity).  When the
latter is varied it affects tunneling exponentially and the
interaction coefficients only as a power law. In what follows, we
therefore take $U$ and $g$ as constant and consider a phase diagram in
the coordinates $t$ and $\mu$.

There are two limits when construction of the phase
diagram of (\ref{Harray}) is simple. When $g=0$
there is no energetic penalty for having  high
angular momentum (non zero spin) states of quantum rotors
in (\ref{Harray}) and spin symmetry is broken.
For integer filling factors we have 
spin nematic insulator phases (NI)
for small $t$ and a superfluid polar condensate phase (PC)
for larger values of $t$. 
For fractional filling factors
the system is always superfluid.
The other simple case is when $g$ is large and states with  odd number
of atoms in individual wells are energetically costly: 
they are not allowed  to have $L=0$ state of a rotor due to 
parity constraint
and have a higher rotational energy.  It is
energetically favorable to have  even numbers of atoms in each well
arranged into singlet combinations, which can be interpreted as 
binding of atoms into singlet pairs with a pair binding energy
$
E_p(Q)=2E(Q+1)-E(Q+2)-E(Q)= g 
$.
In the limit when $g \rightarrow \infty$ 
the crystal phases are  possible for even numbers of atoms
per well only and correspond to spin singlet insulators (SSI). 
The superfluid phase is a condensate of singlet pairs (SSC),
in which tunneling of individual atoms between the wells is
suppressed and only singlet pairs 
are delocalized.
The origin of pairing in this case is not the attraction between
individual bosons, 
but a singlet formation on the
scale of individual wells.  This  is reminiscent of
the ``attraction from repulsion'' mechanism
of electron pairing proposed by Chakravarty
and Kivelson for 
high Tc cuprates, $C_{60}$, and polyacenes \cite{Chakravarty00}.
In the case of finite $g$ we expect  Mott crystal phases  for
all integer filling factors. For even $N$'s and small
$t$ the rotors kinetic energy dominates and we have singlet crystal
phases (SSI). When $t$ is increased the system goes to a spin nematic 
insulator (NI) phase that has admixture of $L \neq 0$ in the ground state. 
For odd $N$'s when all  charge fluctuations are frozen
there is always at least $L=1$ on each site, so we expect the system
to be analogous to spin 1 lattice model with a broken spin symmetry.
The superfluid phase will be
spin singlet fluid of pairs (SSC) for small $t$, and a polar condensate
(PC) for larger $t$'s.

Simple Josephson type experiment may be used to distinguish
superfluid phases suggested above.
We imagine
spinor atoms in an optical lattice in the presense of gravity. Gravity
acts as voltage in a conventional Josephson junction arrays
and gives rise to Josephson oscillations \cite{Kasevich}. 
Josephson oscillations may be measured selectively in spin (1), 
or measuring all the particles
regardless of their spin (2). In the case of single atom
condensate (PC) we expect  oscillations for 
both kinds of measurements
with frequent $\hbar \omega_J = m g \lambda/2$. A singlet pair
condensate will only give oscillations in channel (2)
with frequency  $2 \hbar \omega_J =  m g \lambda$. Insulating phases
will not show any Josephson oscillations, and nematic 
vs spin singlet may be distinguished further through different correlations
for particles of different spin.
\begin{figure*}[h]
\centerline{\epsfxsize=7.0cm 
\epsfbox{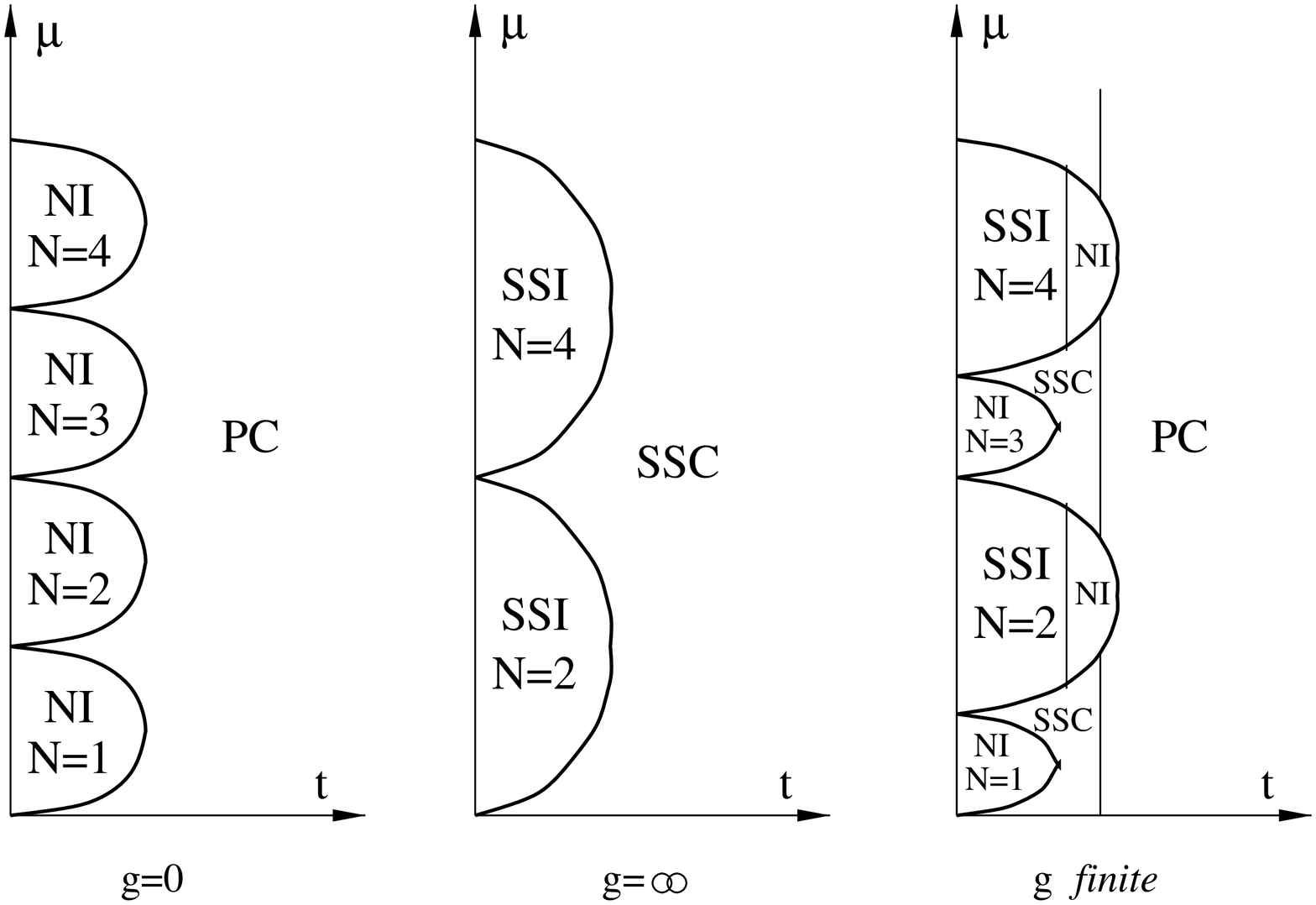}}
\caption{Phase diagram of the Hamiltonian (\ref{Harray})
for different values of parameter $g$.}
\label{phase_diagram}
\end{figure*}

Order parameters for various phases 
suggested above and shown on Figure  \ref{phase_diagram}
may be conveniently discussed using a Euclidean space-time action.
We implement a constraint $N_i+L_{i}=$even 
using a projection operator
$
P_i = \frac{1}{2} \sum_{\sigma_i=\pm1} e^{i\frac{\pi}{2}(1-\sigma_i)(N_i+L_{i})}
$ \cite{Senthil99}.
To calculate the partition function 
$
Z= Tr[ e^{-\beta {\cal H}}P]
$
we divide imaginary time $\beta$ into $M$ slices of length $\epsilon=\beta/M$
and after a few standard manipulations
that include Hubbard-Stratanovich transformation
of the last term in (\ref{Harray}) we find
\begin{eqnarray}
S= - \sum_{rr'} J^c_{rr'} \sigma_{rr'} cos \phi_{rr'}
   - \sum_{rr'} J^{2c}_{rr'}  cos (2 \phi_{rr'})
\nonumber\\
   - \sum_{rr'} J^s_{rr'} \sigma_{rr'} {\bf n}_r {\bf n}_{r'} 
  - \sum_{rr'} J^{2s}_{rr'}  Q^{ab}_r Q^{ab}_{r'} 
\label{Sf}
\end{eqnarray}
Here summation goes over sites $r=(i,\tau)$ in the space-time lattice,
$\phi_r$ is the phase variable conjugate to $N_r$, 
$\phi_{rr'}=\phi_r-\phi_{r'}$,
$Q^{ab}_r = n^a_r
n^b_r - \delta^{ab}/3$ is a nematic order
parameter, and $\sigma_{rr'}= \pm 1$ is an Ising field 
that lives on the links rather than sites. 
The coupling constants
are $J^c_{r,r \pm \hat{\tau}}=(\epsilon u)^{-1}$,
 $J^s_{r,r \pm \hat{\tau}}=(\epsilon g)^{-1}$, 
 $J^{c,s}_{r,r \pm \{\hat{x},\hat{y},\hat{z}\}}=\epsilon J\, |\chi|$,
  $J^{2c}_{r,r \pm \hat{\tau}}=J^{2s}_{r,r \pm \hat{\tau}}=0$,
 $J^{2c,2s}_{r,r \pm \{\hat{x},\hat{y},\hat{z}\}}=-\epsilon J /4$,
with $\chi$ being the saddle point value of the Hubbard-Stratanovich
field.
The usual periodic boundary conditions for $\phi$, ${\bf n}$, 
and $\sigma$ are
assumed at  $\tau=0$ and $\tau=\beta$.
In writing (\ref{Sf}) we omitted the Berry phase terms
that are important for quantitative calculations of the phase diagram
but not for the symmetry arguments discussed below.

Action (\ref{Sf}) has a  $Z_2$ gauge symmetry
$
\phi_r \rightarrow \phi_r +\frac{\pi}{2}(1-\epsilon_r)
$, $  
{\bf n}_r \rightarrow  \epsilon_r\, {\bf n}_r 
%\theta_r \rightarrow \theta_r +\frac{\pi}{2}(1-\epsilon_r)
$, $  
\sigma_{r,r+\hat{a}} \rightarrow \epsilon_r\,\, \sigma_{r,r+\hat{a}} \,\,
\epsilon_{r+\hat{a}}
$, 
where $\epsilon_r = \pm1$.
Such a symmetry has been pointed out earlier
in \cite{Zhou99}, where it was observed that the physical order
parameter is a complex vector ${\bf d} = e^{i \phi} {\bf n}$ 
(not $\phi$ and $ {\bf n}$ separately) that has a symmetry of
$e^{i \phi} \rightarrow - e^{i \phi}$, $ {\bf n}  \rightarrow - {\bf n}$.
Equation (\ref{Sf})
is the simplest action
consistent with the charge $U(1)$, spin $SO(3)$, and gauge $Z_2$
symmetries of the model.
Another term  allowed by the $Z_2$ symmetry 
is the analogue of the Maxwell terms for the lattice gauge
models
$
S_{\sigma} = -K \sum_\Box \prod_\Box \sigma_{ij}
$,
where summation goes over plaquettes in $d+1$ dimensional lattice.
This term may be generated by integrating out the high energy
degrees of freedom.

The existence of the local symmetry imposes important
constraints on the possible order parameters and symmetry breaking
states of the system. Only order parameters that are gauge invariant
may acquire expectation values. For example, the expectation values
$\langle {\bf n} \rangle \neq 0 $ or $\langle e^{i \phi} \rangle \neq 0
$ are not allowed in the models described by (\ref{Sf}), since ${\bf
n}_i$ and $e^{i \phi_i}$ are not invariant under $Z_2$ gauge
transformations.  Physically this restriction arises from the fact
that the wavefunction $\Psi({\bf n})$ should be an even or odd
function of ${\bf n}$ for $N$ even or odd respectively.  Below we
consider several examples of the order parameters that are gauge
invariant and  review broken symmetry phases
that they lead to (see Figure \ref{phase_diagram}).
Polar BEC phase (PC) has $
\langle {\bf d}  \rangle = 
\langle  e^{i \phi} \,\,{\bf n} \rangle \neq 0 
$. It breaks both
charge and spin symmetries and is 
the phase where  atoms are condensed directly.
Nematic Insulator phase (NI) has  $\langle Q_{ab} \rangle 
= \langle n_a n_b -  \delta_{ab}/3 \rangle \neq 0 
$ and breaks only the spin
SO(3) symmetry. This is  a phase where
not the atoms but certain particle-hole composites are condensed.
Spin singlet condensate (SSC)
$\langle e^{ i 2 \phi}   \rangle \neq 0 
$  breaks the 
$U(1)$ symmetry but not $SO(3)$. It corresponds
to a condensate
of singlet pairs of atoms.
Strong coupling pairing phase (SCP) is a phase where both
$\langle e^{ i 2 \phi}   \rangle \neq 0 $
and  $\langle Q_{ab} \rangle \neq 0$.
This phase breaks both $U(1)$ and $SO(3)$ symmetries, however, it is
fundamentally different from the polar condensate phase. 
Transition
between phases PC and SCP should be Ising type transition.
Similar phase has been recently suggested in the context of bilayer 
quantum Hall systems and called ``strong coupling pairing phase''
\cite{Kim00}.
Finally, spin singlet insulator (SSI) is a phase where  no field has 
expectation value, so  neither U(1) nor SO(3) symmetries are broken. 
We note that for a finite number
of traps, when spontaneous symmetry breaking is not possible,
one can observe signatures of the transitions
by measuring fluctuations in the number of atoms,
atom pairs, and spins \cite{Kasevich}.

It is interesting to point out that phases NI, SSC, and SCP
correspond to the fragmented BEC discussed by 
P. Nozieres and D. Saint James.
The part emphasized in \cite{Nozieres82} is that
fragmented state has no macroscopic population of k=0 state for the
bosons. Another interesting aspect of these states is that they
achieve fractionalization of topological objects.
The unfragmented PC phase has topological excitations that
are composites of 1/2 vortex in the charge sector and
$\pi$-disclination in the spin sector. 
Here  1/2 charge vortex is a
topological defect around which the phase $\phi$ winds by $\pi$ and
$e^{i \phi}$ changes sign.  $\pi$-disclination in the spin sector is
introduced as a topological defect around which ${\bf n}$ changes
sign (vortex like objects in ${\bf n}$ are called merons, 
so $\pi$-disclination is 1/2 of a meron).
Both 1/2 vortex and $\pi$-disclination 
are pointlike objects in two dimensions and lines
in three dimensions. No individual 1/2 charge vortices or
$\pi$-disclinations are allowed in the PC phase.
The spin and charge parts of
the vortex are always glued together.  
A condensate of singlet pairs only, the SSC phase, has 1/2 charge
vortices as separate excitations, but no spin vortices; and the
nematic phase NI has $\pi$ disclinations and no charge vortices.  A
condensate of S=0 pairs and S=2 pairs, a SCP phase, will have 1/2 charge
vortices and $\pi$ disclinations as separate excitations.

Establishing the nature of the broken symmetry does not fully
characterize models with gauge symmetry. 
Another important aspect of the system described by (\ref{Sf}) is the
possibility of confining and deconfining phases of the $Z_2$ gauge
theory in dimension $d+1 \geq 3$ \cite{Senthil99,Fradkin79,Sachdev91}.
%As discussed in \cite{Senthil99,Fradkin79} 
%the confinement deconfinement
%transition exists even in the presense of matter fields and leads to
%an additional distinction within the states with and without broken
%continuous symmetry. 
Difference between the confining and deconfining
phases relies on the concept of topological order \cite{Wen}
and  has important implications on the nature of excitations
in the system. Topological order
may coexist with any true long range order, so any of the phases
reviewed earlier may be confining or deconfining.
For the pure  $Z_2$
gauge models
the confining and the deconfining phases 
may be distinguished using Wilson loops, which  are
characterized by an area law for the confining phase and by a
perimeter law for the deconfining phase.  
With matter field  present Wilson loops may no longer be
used to discriminate between the two phases \cite{Fradkin79}, however
distinction between them survives, and there is 
a phase transition between the two. A simple way 
to understand the transition is to
think of it as condensation of visons, topological excitations of the
$Z_2$ gauge theory that describe frustrated plaquettes $\prod_{\Box}
\sigma_{rr'}=-1$. Particles that carry a $Z_2$ charge are frustrated
when traveling around a vison, so when visons are condensed such
particles may not propagate coherently and we have a confining phase.
When visons are not condensed, they are finite energy topological
excitations and we have a deconfined phase.
Presense of finite energy visons leads to an additional degeneracy of
the gauge models in a deconfined state on manifolds with non-trivial
degeneracy. Systems with $Z_2$ gauge symmetries in condensed matter
have already been discussed by Lammert {\it et.al.} \cite{Lammert93}
in connection with liquid crystals and Senthil and Fisher in
connection with high temperature superconductors \cite{Senthil99}.
For our system the discussion above implies that all the broken symmetry
states , as well as a state SSI, where neither U(1)
nor SO(3) symmetries are broken, come in two varieties: 
confining and deconfining.
Detailed comparison between confining and deconfining
versions of various will be given in \cite{long}. Here
we only note that some of the most striking implications of 
deconfinement appear for the
SSI phase. The deconfining SSI$^*$ phase allows charge $b=e^{i\phi}$ and
spin ${\bf n}$ carrying excitations propagate independently
\cite{disclaimer2}.

Another perspective on the nature of deconfining phases comes from
considering phase transitions as condensation of topological defects.
Discussion will be given in the case of two spatial dimensions,
where condensation of point-like topological objects (vortices,
$\pi$-disclinations, merons, and etc.) drive transitions into the
phases of restored symmetry. We expect, however, that most of the
conclusions reached for $d=2$ will also hold in $d=3$.
Let us consider a transition between
the PC phase  that breaks charge and spin symmetry 
and NI phase that breaks only the spin symmetry. 
In the phase PC we have topological objects that are 
1/2 vortex $\pi$-disclination composites. Half-vortex and half-meron 
parts 
may come with different signs, so we have 4 types of 
these composites ($\pm$1/2, $\pm$1/2), where
the first and the second numbers stand for the charge  and the spin 
parts respectively.
To go to a phase that has
broken spin symmetry and restored charge symmetry 
one expects that we need
to condense topological objects in the charge sector,
with topological objects in the spin sector remaining
at finite energy. This can be achieved by condensing  composites
$(1/2,1/2) + (1/2,-1/2) = (1,0)$
and
$(-1/2,1/2) + (-1/2,-1/2) = (-1,0)$, but 
having individual topological defects $(\pm1/2,\pm1/2)$ 
gapped.
Such transition would correspond to condensation of integer vortices 
and would be equivalent 
to order-disorder
transition for the charge order parameter $e^{i \phi}$
in the usual Bose-Hubbard model.
An important observation is that the insulating phase
obtained this way is a fractionalized phase NI$^*$! It does not have
a condensate of minimal topological object in the charge sector
(1/2 vortex), but only a condensate of integer vortices.
So, only the $PC -  NI^*$ transition
may be a continuous second order
transition, but  $PC-  NI$ must be a first order
transition. The existence of  fractionalized phases
may, therefore, be established through  study of phase transitions
\cite{Lammert93}.
We note, however, that starting
from a strong coupling pairing phase SCP (that
has unbound 1/2- vortices and $\pi$-disclinations) 
we will have $SCP -  NI$ as a second order transition.

To summarize, we reviewed several novel phenomena that will take
place for $F=1$ Bose gases with antiferromagnetic interactions. 
We showed that in optical lattices they will
have several distinct superfluid and Mott insulator 
phases depending on the density and parameters
of the system.
Another possibility suggested in this
paper is a class of fractionalized phases of spinor BEC, characterized
by the topological order that may coexist with any true long range
order in the system. Quantitative analysis
of experimental systems will be given in subsequent 
publications \cite{long}.
Discussion in this paper may be generalized to systems with
higher hyperfine spins. We also expect some of the results to 
be applicable to triplet superconductors.

We thank  M. Fisher, H.Y. Kee,  Y.B. Kim, M. Lukin, C. Nayak, S. Sachdev, 
J. Sethna, T. Senthil, H. T. Stoof, P. Wiegnman
and particularly F.D. Haldane for useful discussions. 
ED is supported by the Harvard Society of Fellows.


\begin{thebibliography}{99}
\bibitem{disclaimerNlarge}
Hamiltonian (\ref{Harray}) is derived in the limit $\bar{N}
>>1$. Qualitative picture derived from its analysis should, however,
apply as long as $\bar{N} \geq 1$.
\bibitem{disclaimer2}
Even in a confining SSI
phase, two spin 1 excitations carrying charges +1 and -1 may combine
to produce a charge neutral spin 1 quasiparticle.
Nevertheless, phase SSI$^*$ is a
direct analogue of the spin-charge separated phases suggested for the
high Tc cuprates, and the true distinction between the SSI and SSI$^*$ phases
is topological. By restricting the
spin symmetry  to SO(2) one can make the statement of spin charge
separation precise.
\end{thebibliography}
\end{document}